# Recent progress of RF cavity study at Mucool Test Area

**Katsuya Yonehara on behalf of MTA working group**

Fermilab, MS 221, P.O. Box 500, Batavia, IL 60510, USA

yonehara@fnal.gov

**Abstract**. In order to develop an RF cavity that is applicable for a muon beam cooling channel, a new facility, called Mucool Test Area (MTA) has been built at Fermilab. MTA is a unique facility whose purpose is to test RF cavities in various conditions. There are 201 and 805 MHz high power sources, a 4-Tesla solenoid magnet, a cryogenic system including a Helium liquifier, an explosion proof apparatus to operate gaseous/liquid Hydrogen, and a beam transport line to send an intense H$^-$ beam from the Fermilab Linac accelerator to the MTA hall. Recent activities at MTA will be discussed in this document.

## 1. Introduction

A muon collider is an attractive possibility for the next generation of colliders at the energy frontier. The muon provides clean collisions and due to its mass, there is ignorable energy loss from bremsstrahlung compared to an electron. Since the muon has a short lifetime, one of greatest challenges in a muon acceleration is how quick a muon beam is accelerated to achieve the desired beam energy. This requires muon phase space cooling in the acceleration process. Due to the characteristics of muon, an ionization cooling method is the most effective way. From ionization cooling simulations, the muon beam six-dimensional phase space can be cooled more than $10^6$ of magnitude down within the muon lifetime [1,2]. However, a muon ionization cooling channel has never been built since the beam lattice is more complicated than a conventional beam channel. Because a beam dynamics in the ionization cooling process is perturbed by energy loss and stochastic interactions the beam lattice of cooling channel needs to have a larger acceptance than usual. The Muon International Collaboration Experiment (MICE) has begun to investigate the ionization cooling theory and its practicality [3].

A conceptual muon cooling channel consists of a focusing magnet, ionization cooling absorber, and RF cavity. Incident muon beam in the channel is minimized a beta function at the absorber by the magnet, that is maximized the amplitude of transverse momentum of the beam in the absorber. The amplitude of momentum is isotropically reduced by the ionization energy loss process, and an RF cavity restores the energy lost in longitudinal direction. As a result, only the transverse muon beam phase space is cooled. The cooling absorber should be a low z material to minimize a multiple scattering effect. For this reason, gaseous or liquid Hydrogen, solid Beryllium, or solid LiH prefers to be used as an ionization cooling absorber. The energy restore RF cavity needs to be located in a strong magnetic field to make a compact ionization cooling channel. However, the available RF field gradient is strongly limited by the strength of an external magnetic field. A dark current in the cavity is focused by the external magnetic field and accelerated by an RF electric field. This creates larger damage on the cavity surface by dumping its energy to smaller spot and induces RF breakdown at lower electric field in stronger magnetic fields. The goal electric field is 25 MV/m at 3 Tesla. But, the present available electric field is about a half of the goal value [4]. In order to investigate this problem, we have built a new test facility, Mucool Test Area (MTA) and have tested several conceptually new RF cavities. This document will introduce a unique feature of MTA and summarize a recent result of RF cavity tests.

**2. MTA**
A main purpose of MTA is to observe a characteristic of an RF cavity under the same condition as a muon ionization cooling channel. MTA has many unique instruments to accomplish this purpose. There are 201 and 805 MHz high power RF sources and waveguides. Peak powers of RF sources are 4.5 and 12 MW for 201 and 805 MHz, respectively. There is an RF switch on the 805 MHz waveguide to share the RF power into two RF cavities. There is a 4-Tesla superconducting solenoid magnet that has a 44 cm bore hole to assemble an 805 MHz cavity in the magnet. There is a liquid helium recovery system that has a large cooling power, 250 W enough to cool down the magnet as well as other cryogenic system. The apparatus in the MTA experimental hall is explosion proofed to operate gaseous and liquid Hydrogen. The MTA experimental hall is kept cleaning (Class 100) to handle a precision instrument. Recently, the commission of MTA beam line has been completed [5]. Now a 400 MeV full Linac H$^-$ beam ($7\times10^{12}$ H/pulse) is used for the RF cavity test although the maximum beam repetition rate is 60 pulses per hour due to the limit of radiation hazard.

**3. Test special RF cavity**

3.1. 201 MHz RF cavity
A key issue to improve the RF cavity performance in a strong magnetic field is how to remove a dark current or reduce its intensity in the cavity. A possible source of dark current is a surface emission electron. This mechanism is well known as a Fouler-Nordheim model. According to the model, a probability of electron emission depends on a field enhancement factor related with a surface roughness, a work function of a cavity material, and the strength of an applied electric field. By treating a cavity wall surface with a special rinsing technique that is used for a superconducting RF cavity, the surface roughness of a normal conducting cavity becomes drastically small. Besides, using a Beryllium RF window that has a high work function makes lower probability of surface emission of electron than Copper. The special surface-treated 201 MHz pillbox cavity with a Beryllium RF window has been tested under a magnetic field by using a fringe field of the 4-Tesla solenoid. The observed electric fields were 21 and 12 MV/m at 0 and 0.75 Tesla at the center of the cavity. The cavity will be tested with a real ionization cooling magnet to measure the available electric field in 2012.

3.2. 805 MHz button RF cavity
In order to investigate the property of cavity material, an 805 MHz cavity that has a replaceable "button" electrode has been tested under a strong magnetic field. We have tested Copper, Molybdenum, Tungsten, and Tin coated Copper materials. But, none of these reached to the goal value in a strong magnetic field [6]. Unfortunately, the RF power line needs to access from longitudinal direction of the cavity in the present cavity configuration. This structure makes a high electric field spot near the RF power port. In fact, there were many pits that were made during the breakdown process around the port. This can cause an uncertainty of the measurement to find the available RF gradient. The cavity has been refurbished to eliminate such an unwanted hot spot. We plan to repeat the experiment in fall 2011.

3.3. 805 MHz box RF cavity
Another unique 805 MHz cavity test has been made. It is a box cavity to configure the RF electric field direction perpendicular to the external magnetic field. In this field configuration, a dark current is diverged from a high electric field region to the low one by an E×B force [7]. As a result, a surface emission electron is not accelerated by the electric field and absorbed on the cavity wall without heating up the surface. Experiment has been made and the observed available electric fields were 50 and 33 MV/m in 0 and 3 Tesla at the E×B angle 90 degrees, respectively. However, an allowable angle between electric and magnetic fields is only 90 ± 2 degrees. It requires a special shape RF cavity that has a half of shunt impedance as a normal pillbox cavity [8]. It will be expensive.

3.4. Gas-filled 805 MHz RF cavity

A gas filled 805 MHz RF cavity contains a dense buffer gas in the cavity to stop electron acceleration by an RF field. The gas even avoids focusing of the dark current in a strong magnetic field. Besides, the gas can be used as the ionization cooling absorber when gaseous hydrogen is filled in the cavity. A high pressure hydrogen gas filled RF cavity has been demonstrated in a strong magnetic field and no RF field degradation has been observed [9]. A crucial issue on this cavity is a beam-induced plasma loading effect. Ionized electrons in the cavity, which is produced by an interaction between an incident charged particle and a dense Hydrogen gas, can consume a stored RF power. Even ionized electrons may induce the RF breakdown. The beam test has been made at MTA to measure the beam-induced plasma loading effect. First of all, no RF breakdown has been observed. The observed RF power consumption rate is good agreement with the expected value that comes from a plasma dynamics model in a DC field. For example, an RF power consumption of single electron is $5 \times 10^{-17}$ Joules/cm/RF cycle/electron at applied RF electric field 20 MV/m. An electronegative dopant gas effect has also been tested. A 0.01 % of SF6 was doped in a pure hydrogen gas. As a result, the electron capture rate in the doped hydrogen gas was two orders of magnitudes better than that in a pure hydrogen gas. Figure 1 shows the plasma loading effect in a pure Hydrogen gas (red) and Hydrogen gas with 0.01 % SF6 dopant gas (blue). It is clear that the plasma loading is significantly mitigated by SF6. The detail analysis is under study.

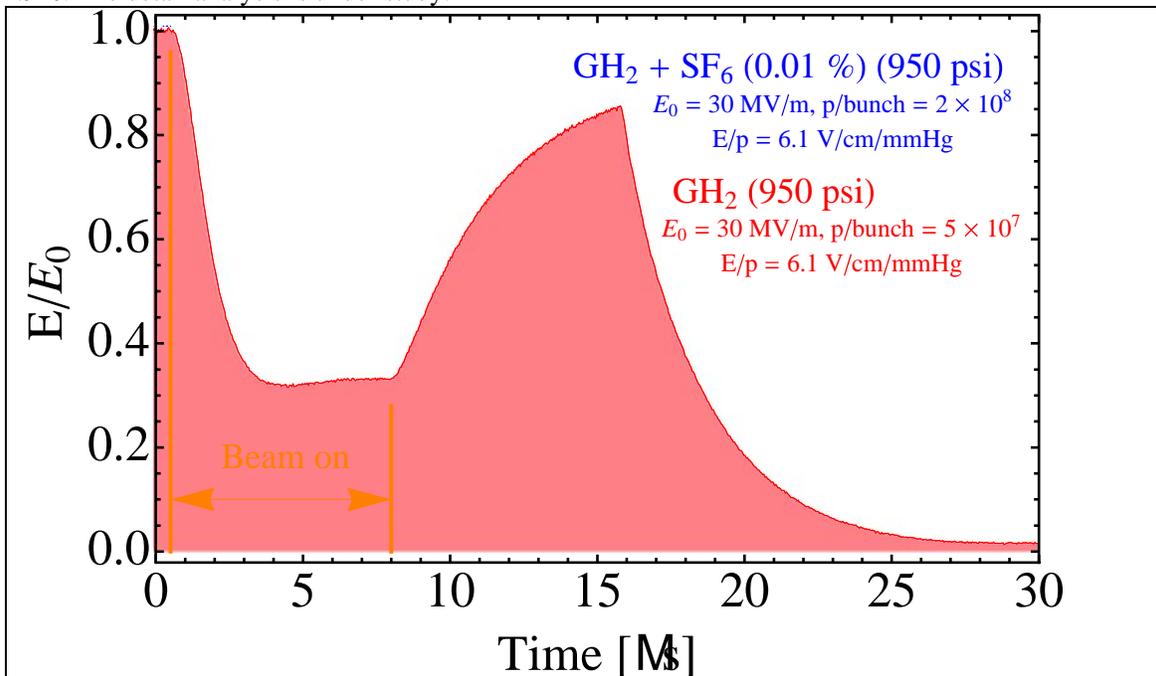

Figure 1: Observed RF pickup signals in a pure hydrogen gas and hydrogen gas with 0.01 % SF6 dopant gas.

4. Conclusion

Within the first five years of the MTA, development of RF technology for an ionization cooling channel is progressing at a rapid pace. Gas-filled RF cavities have been demonstrated to drastically mitigate the effects of RF breakdown, and are therefore a good candidate for a cooling channel. A total RF power consumption of the plasma loading in a gas-filled RF cavity with a real muon beam for a collider can be estimated from the experimental result. The number of ionized electrons in one muon beam bunch will be $10^{14}$~$10^{16}$ electrons. Thus, from the observed RF power consumption per single electron ($5 \times 10^{-17}$ Joules/RF cycle/electron/cm), a total RF power consumption is 0.005~0.5 Joules/RF cycle/cm. It corresponds to 0.1~10 % of stored energy in a 200 MHz pillbox cavity in a unit length. These values can be overestimated at dense plasma condition in a real cooling channel since this simple power loss estimation does not include the recombination process. We need more beam test.

We also demonstrated that the beam plasma density was significantly reduced by using a small amount of electronegative dopant gas. 0.01 % of $SF_6$ dopant reduces factor 100 of the electron density. However, $SF_6$ gas may not be practical or the ionization cooling channel since the channel will be operated at cryogenic temperature. Oxygen will be a good candidate of capturing electrons since the lowest flammable level (LFL) of Oxygen molecule in Hydrogen gas is 7 % that is two orders of magnitude higher than the required amount of doping rate to remove the electron swarm. Researching electronegative gas should be made.

Once we find the possible RF cavity for an ionization cooling channel, we will integrate all muon beam cooling elements including with an infrastructure and demonstrate an entire system in MTA.